\documentclass[twocolumn,showpacs,aps,prl,superscriptaddress,letterpaper,floatfix]{revtex4-1}
\usepackage{graphicx}
\usepackage{dcolumn}
\usepackage{amsmath}
\usepackage[dvips]{color}
\usepackage{bm}

\RequirePackage{xspace}
\def\epem {\ensuremath{e^+e^-}\xspace}
\newcommand{\gev}{\ensuremath{\mathrm{\,Ge\kern -0.1em V}}\xspace}
\newcommand{\gevcc}{\ensuremath{{\mathrm{\,Ge\kern -0.1em V\!/}c^2}}\xspace}
\newcommand{\tev}{\ensuremath{\mathrm{\,Te\kern -0.1em V}}\xspace}
\def\km   {\ensuremath{{\rm \,km}}\xspace}
\def\cm   {\ensuremath{{\rm \,cm}}\xspace}
\def\mm   {\ensuremath{{\rm \,mm}}\xspace}
\def\mum  {\ensuremath{{\,\mu\rm m}}\xspace}
\def\nm   {\ensuremath{{\rm \,nm}}\xspace}
\def\cms  {\ensuremath{{\rm \,cm}^{-2} {\rm s}^{-1}}\xspace}
\newcommand{\lum} {\ensuremath{\mathcal{L}}\xspace}

\begin{document}

\preprint{BINP 2012-5}

\title{ \boldmath Restriction on the energy and luminosity of \epem
  storage rings due to beamstrahlung}

\author{V.~I.~Telnov}
\email{telnov@inp.nsk.su}
\affiliation{ Budker Institute of Nuclear Physics SB RAS, 630090, Novosibirsk, Russia \\
Novosibirsk State University, 630090, Novosibirsk, Russia}

\date{29 March 2012}

\begin{abstract}

The role of beamstrahlung in high-energy \epem storage-ring colliders (SRCs)
is examined. Particle loss due to the emission of single energetic
beamstrahlung photons is shown to impose a fundamental limit
on SRC luminosities at energies
$2 E_0 \gtrsim 140 \gev$ for head-on collisions and $2 E_0 \gtrsim 40 \gev$
for crab-waist collisions.
With beamstrahlung taken into account, we explore the viability of SRCs in the
$2E_0=240$--500 \gev range, which is of interest
in the precision study of the Higgs boson.
At $2E_0=240 \gev$, SRCs are found to be competitive with linear colliders;
however, at $2E_0=400$--500\gev, the attainable SRC luminosity would be a factor
15--25 smaller than desired.

\end{abstract}


\maketitle

The ATLAS and CMS experiments at the LHC recently
reported~\cite{higgsATLAS,higgsCMS} an excess
of events at $M = 125\gevcc$, which may be evidence for the
long-sought Higgs boson. The precision study of the Higgs boson's
properties  would require the construction of an
energy- and luminosity-frontier \epem collider~\cite{Aarons}.

The $2E_0=209 \gev$ LEP collider at CERN is generally considered to
have been the last energy-frontier \epem storage-ring collider
(SRC) due to synchrotron-radiation energy losses, which are proportional to
$E_0^4/R$. Linear \epem colliders (LC) are free from this
limitation and allow multi-\tev energies to be reached.  Two LC
projects are in advanced stages of development: the $2E_0=500 \gev$
ILC~\cite{ILC} and the $2E_0=500$--3000\gev CLIC~\cite{CLIC}.

Nevertheless, several proposals~\cite{Zim,Oide} for a $2E_0=240 \gev$
SRC for the study of the Higgs boson in $\epem \to HZ$ have recently
been put forward~\cite{Sen}.  Lower cost and reliance on firmly
established technologies are cited as these projects' advantages over
an LC.  Moreover, it has been proposed that a $2E_0=240 \gev$ SRC can
provide superior luminosity, and that the ``crab waist'' collision
scheme~\cite{crab,SuperB} allows them to exceed the ILC and CLIC
luminosities even at $2E_0=400$--500 \gev.  Parameters of the recently
proposed SRCs are summarized in Table~\ref{Table1}.

The present paper examines the role of {\sl beamstrahlung}, i.e.,
synchrotron radiation in the field of the opposing beam, in
high-energy \epem SRCs.  First discussed in~\cite{Rees}, beamstrahlung
has been well-studied only in the LC case~\cite{Chen}.  As we shall
see, at energy-frontier \epem SRCs beamstrahlung determines the beam
lifetime through the emission of single photons in the tail of the
beamstrahlung spectra, thus severely limiting the luminosity.
\begin{table*}[!hbt]
\caption{Parameters of LEP and several recently proposed storage-ring
colliders~\cite{Zim,Oide}.
``STR'' refers to ``SuperTRISTAN''~\cite{Oide}.
Use of the crab-waist collision
scheme~\cite{crab,SuperB} is denoted by ``cr-w''.
 The luminosities and the numbers of bunches for all projects
 are normalized to the total synchrotron-radiation power of 100 MW.
Beamstrahlung-related quantities
derived in this paper are listed below the double horizontal line.}
{\renewcommand{\arraystretch}{0.88}
\begin{ruledtabular}
\begin{tabular}{l c c c c c c c c c c c}  \\[-2.5mm]
& LEP & LEP3 & DLEP & STR1 & STR2 & STR3 &  STR4 &  STR5 &  STR6 \\[-1mm]
&&&&&&cr-w&cr-w&cr-w&cr-w \\
\hline \\[-2.2mm]
$2E_0$, \gev      & 209 & 240 & 240 & 240 & 240 & 240 & 400 & 400 & 500   \\
Circumference,\km & 27& 27 & 53 & 40 & 60 & 40 & 40 & 60 & 80 \\
Beam current, mA  & 4 & 7.2 & 14.4 & 14.5 & 23 & 14.7 & 1.5 & 2.7 & 1.55 \\
Bunches/beam & 4  & 3 & 60 & 20 & 49 & 15 & 1 & 1.4 & 2.2 \\
$N,\;10^{11}$          & 5.8 & 13.5 & 2.6 & 6 & 6 & 8.3 & 12.5 & 25.& 11.7  \\
$\sigma_{z}$,\mm    & 16  & 3   & 1.5 & 3   &  3 & 1.9 & 1.3 & 1.4 & 1.9  \\
$\varepsilon_x$/$\varepsilon_y$,\nm   & 48/0.25   & 20/0.15   & 5/0.05  & 23.3/0.09& 24.6/0.09& 3/0.011   & 2/0.011  & 3.2/0.017& 3.4/0.013\\
$\beta_x$/$\beta_y$,\mm      & 1500/50 & 150/1.2 & 200/2 & 80/2.5  & 80/2.5  & 26/0.25  & 20/0.2 & 30/0.32 & 34/0.26 \\
$\sigma_x$/$\sigma_y$,\mum & 270/3.5 & 54/0.42   & 32/0.32 & 43/0.47  & 44/0.47  & 8.8/0.05 & 6.3/0.047 & 9.8/0.074 & 10.7/0.06 \\
SR power, MW       & 22  & 100  & 100 & 100  & 100  & 100 & 100 & 100 & 100 \\
Energy loss/turn,\gev & 3.4 & 7 & 3.47 & 3.42 & 2.15 & 3.42 & 33.9 & 18.5 & 32.45 \\
\lum, $10^{34}\cms$ & 0.013 & 1.3 & 1.6 & 1.7 & 2.7 & 17.6 & 4 & 7 & 2.2 \\
\hline \hline \\[-2.6mm]
$E_\mathrm{c,max}/E_0$,$\, 10^{-3}$ & 0.09 & 6.3 & 4.2 & 3.5 & 3.4 & 38 & 194 & 232 & 91 \\
$n_{\gamma}$/electron    & 0.09 & 1.1 & 0.37 & 0.61 & 0.6 & 4.2 & 8.7&  11.3 & 4.8 \\
lifetime(SR@IP), s (Eq.~{\ref{5}}) & $\sim \infty$ & 0.02 & 0.3 & 0.2 & 0.4 & 0.005 & 0.001 & 0.0005 & 0.005 \\
\hline \\[-2.6mm]
$\lum_\mathrm{corr}$, $10^{34}\cms$ & 0.013 & 0.2 & 0.4 & 0.5 & 0.8 & 0.46 & 0.02 & 0.03 & 0.024 \\
\end{tabular}
\end{ruledtabular}
\vspace{-4mm}
}
\label{Table1}
\end{table*}

At SRCs the particles that lose a certain fraction of their energy in
a beam collision leave the beam; this fraction $\eta$ is typically
around 0.01 (0.012 at LEP) and is known as the ring's energy
acceptance.  Beamstrahlung was negligible at all previously built
SRCs. Its importance considerably increases with energy.
Table~\ref{Table1} lists the beamstrahlung characteristics of the
newly proposed SRCs assuming a 1\% energy acceptance: the critical
photon energy for the maximum beam field $E_\mathrm{c,max}$, the
average number of beamstrahlung photons per electron per beam crossing
$n_{\gamma}$, and the beamstrahlung-driven beam lifetime.  Please note
that once beamstrahlung is taken into account, the beam lifetime drops
to unacceptable values, from a fraction of a second to as low as a few
revolution periods.

At the SRCs considered in Table~\ref{Table1}, the beam lifetime due to
the unavoidable radiative Bhabha scattering is 10 minutes or longer.
One would therefore want the beam lifetime due to beamstrahlung to be
at least 30 minutes. The simplest (but not optimum) way to suppress
beamstrahlung is to decrease the number of particles per bunch with a
simultaneous increase in the number of colliding bunches.  As
explained below, $E_\mathrm{c,max}$ should be reduced to $\approx
0.001E_0$. Thus, beamstrahlung causes a great drop in luminosity,
especially at crab-waist SRCs: compare the proposed \lum and corrected
(as suggested above) $\lum_\mathrm{corr}$ rows in Table~\ref{Table1}.

 To achieve a reasonable beam lifetime, one must make small
the number of beamstrahlung photons with energies greater than
the threshold energy $E_\mathrm{th}=\eta E_0$ that causes the electron
to leave the beam. These photons belong to the high-energy
tail of the beamstrahlung spectrum and have energies much greater
than the critical energy. It will be  shown below that the
beam lifetime is determined by such single high-energy
beamstrahlung photons, not by the energy spread due to the emission
of multiple low-energy photons.

The critical energy for synchrotron radiation~\cite{pdg}
\begin{equation}
E_\mathrm{c}=\hbar \omega_\mathrm{c} = \hbar\frac{3\gamma^3c}{2\rho},
\label{1}
\end{equation}
where $\rho$ is the bending radius and $\gamma=E_0/mc^2$. The
spectrum of photons per unit length with energy well above the
critical energy~\cite{pdg}
\begin{equation}
\frac{dn}{dx} = \sqrt{\frac{3\pi}{2}}\frac{\alpha \gamma}{2\pi \rho}
\frac{e^{-u}}{\sqrt{u}}du,
\label{2}
\end{equation}
where $u=E_{\gamma}/E_\mathrm{c}$, $\alpha=e^2/\hbar c$.
To evaluate the integral of
this spectrum from the threshold energy $\eta E_0$ to $E_0$ note that
the minimum value of $u \gg 1$, the exponent decreases
rapidly, and so one can integrate only the exponent and use
the minimum value of $u$ outside the exponent.
After integration
and substitution of $\rho$ from Eq.~\ref{1}, we obtain the number of photons
emitted on the collision length $l$ with energy $E_{\gamma} \geq \eta E_0$:
\vspace{-1.5mm}
\begin{equation}
 n_{\gamma}(E_{\gamma}\geq\eta E_0)\approx
\frac{\alpha^2\eta l}{\sqrt{6\pi}r_e\gamma u^{3/2}} e^{-u};
\,\,\,u=\frac{\eta E_0}{E_\mathrm{c}}, \label{3}
\end{equation}
where $r_e=e^2/mc^2$ is the classical radius of the electron.

The regions of the beam where the field strength is the greatest
contribute the most to the emission of the highest-energy photons.
We need to find the critical energy for this field
and the bunch size that yields an acceptable rate of
beamstrahlung particle loss.
The collision length $l \approx \sigma_z/2$ for head-on and
$\approx \beta_y/2$ for crab-waist collisions. In the transverse
direction, we can assume that the electron crosses the region with the
strongest field with a 10\% probability. The average number of
beam collisions $n_\mathrm{col}$ experienced by an electron before it leaves
the beam can be estimated from $0.1n_\mathrm{col}n_{\gamma}=1$, where
$n_{\gamma}$ is given by Eq.~\ref{3}. Thus, $n_\mathrm{col}$ and the beam
lifetime due to beamstrahlung $\tau$
\begin{equation}
n_\mathrm{col} \approx 10 \frac{\sqrt{6\pi}r_e\gamma u^{3/2}}{\alpha^2\eta
l}e^u; \;\;\;\; \tau = n_\mathrm{col} \frac{2\pi R}{c}.
\label{5}
\end{equation}

Assuming $E_0 = 150 \gev$, $l=0.1\cm$, $\eta=0.01$,
and a ring circumference of 50\km,
from Eqs.~\ref{3} and \ref{5} we get
\begin{equation}
u = \eta E_0 / E_\mathrm{c} \approx 8.5; \;\;\;\;
E_\mathrm{c} \approx 0.12\eta E_0 \sim 0.1\eta E_0.
\label{7}
\end{equation}
The accuracy of this expression is quite good for any SRC because
it depends on the values in front of the exponent in Eq.~\ref{5}
only logarithmically.

Let us express the critical energy $E_\mathrm{c}$ via the beam parameters.
In beam collisions, the electrical and magnetic forces are equal
in magnitude
and act on the particles in the oncoming beam in the same direction.
Thus, we can use the effective doubled magnetic field. The maximum
effective field for flat Gaussian beams
$B \approx 2eN/\sigma_x\sigma_z$.
The bending radius
$\rho=pc/eB \approx \gamma
  mc^2/eB=\gamma \sigma_x\sigma_z/2r_e.$
Substituting to Eq.~\ref{1}, we find \vspace{-2mm}
\begin{equation}
\frac{E_\mathrm{c}}{E_0}=\frac{3\gamma {r_e}^2 N}{\alpha \sigma_x \sigma_z}.
\label{10}
 \end{equation}
Combined with Eq.~\ref{7}, this imposes a restriction on
the beam parameters,
\begin{equation}
\frac{N}{\sigma_x \sigma_z} < 0.1 \eta\frac{\alpha}{3\gamma {r_e}^2}.
\label{11}
\end{equation}
{\sl This formula is the basis for the following discussion}.

For Gaussian beams, the average number of beamstrahlung
photons per electron for head-on collisions~\cite{Chen}
$\langle n_{\gamma}\rangle \approx 2.12 N\alpha r_e /\sigma_x$,
their average energy
$\langle E_{\gamma}\rangle \approx 0.31\langle E_\mathrm{c} \rangle$,
and the average critical energy
$\langle E_\mathrm{c} \rangle \approx 0.42E_\mathrm{c, max}$;
hence,
$\langle E_{\gamma}\rangle \approx 0.13 E_\mathrm{c,max} $.
Above, we considered the maximum field, i.e., $E_\mathrm{c}$ was
equal to $E_\mathrm{c, max}$.
Then, for the condition in Eq.~\ref{11} we obtain
\begin{equation}
\langle n_{\gamma} \rangle =\frac{0.07 \eta
  \alpha^2 \sigma_z}{r_e \gamma} \approx \frac{ 0.067
  (\sigma_z/\mbox{mm})}{(E_0/100\gev)}\left(\frac{\eta}{0.01}\right),
\label{ng}
\end{equation} \vspace{-4mm}
\begin{equation}
\langle E_{\gamma} \rangle \approx 0.13\times 0.1\eta E_0 \approx
1.3\times10^{-2}\eta E_0.
\label{eg}
\end{equation}
For crab-waist collisions, $\langle E_{\gamma}\rangle$ is the same
while the interaction length is shorter, $\beta_y$ instead of
$\sigma_z$; therefore, the number of photons is proportionally
smaller.

So,  when $\tau$ is large enough $\tau$ is determined by the rare photons
with energies $\gtrsim 8.5 E_\mathrm{c,max}$, a factor $8.5/0.13=65$
greater than $\langle E_{\gamma}\rangle$.

The beam energy spread due to beamstrahlung can be estimated as
follows. In the general case~\cite{Wied},
\begin{equation}
 \frac{\sigma_E^2}{E_0^2}=\frac{\tau_\mathrm{s}}{4E_0^2}\dot{n}_\gamma\langle
 E_\gamma^2\rangle.
 \end{equation}
In our case, the damping time (due to radiation in bending magnets)
$\tau_\mathrm{s}\approx T_\mathrm{rev} E_0/\Delta E_\mathrm{rev}$,
$\dot{n}_\gamma=\langle n_\gamma \rangle/T_\mathrm{rev}$, and
$\langle E_\gamma^2\rangle \approx 4.3 \langle E_\gamma\rangle^2$~\cite{Sands},
which gives
\begin{equation}
\frac{\sigma_E^2}{E_0^2} \approx \frac{\langle n_\gamma\rangle \langle
 E_\gamma\rangle^2}{E_0\Delta E_\mathrm{rev}}=\frac{1.15\times
 10^{-9}(\sigma_z/\mbox{mm})}{(E_0/100\gev)(\Delta E_\mathrm{rev}/E_0)}\left(\frac{\eta}{0.01}\right)^3,
\label{e-sp-b}
\end{equation}
where $\Delta E_\mathrm{rev}$ is the energy loss per revolution and
$\langle n_\gamma\rangle$ and $\langle E_\gamma\rangle$ are given by
Eqs.~\ref{ng} and \ref{eg}. Taking the typical bunch length
$\sigma_z=5 \mm$, $E_0=120$ GeV, and $\Delta E_\mathrm{rev}/E_0 =0.05$
we get an estimate for the energy spread due to beamstrahlung (under
the condition in Eq.~\ref{11}) $\sigma_E/E_0 \approx 3\times 10^{-4}
(\eta/0.01)^{3/2}$.

The beam energy spread due to synchrotron radiation (SR) in
the bending magnets~\cite{Wied}:
\begin{equation}
 \left( \frac{\sigma_E^2}{E_0^2}\right)_\mathrm{SR}=\frac{55\sqrt{3}}{128\pi \alpha
   J_\mathrm{s}}\frac{mc^2}{E_0} \frac{\Delta E_\mathrm{rev}}{E_0} =
 \frac{0.016}{J_\mathrm{s}E_0(\gev)} \frac{\Delta E_\mathrm{rev}}{E_0},
\end{equation}
where $1<J_\mathrm{s}<2$ is the partition number. For the projects in
Table~\ref{Table1}, $\sigma_E/E_0$ due to SR varies between 0.17\%
and 0.24\%. For $E_0=120\gev$, $\Delta E_\mathrm{rev}/E_0=0.05$,
$J_s=1.5$ one gets $(\sigma_E/E_0)_\mathrm{SR}\approx 2\times 10^{-3}$.
For the given example, the beamstrahlung energy spread becomes larger
than that due to SR in rings at $\eta>0.035$. 

The energy spread due to beamstrahlung contributes to the beam
lifetime (if the lifetime is large enough) when the energy acceptance
$\eta \lesssim 6(\sigma_E/E_0)$; with [\ref{e-sp-b}] taken into account, this
yields $\eta>2.5(\Delta E_\mathrm{rev}/10\gev)/(\sigma_z /\mbox{mm})$.
For the typical $\Delta E_\mathrm{rev}=5\gev$, $\sigma_z=5$ mm, we get
$\eta>0.25$, which is much larger than the realistic storage-ring
energy acceptance $\eta =$ 0.01--0.03.
Therefore, the beam energy spread due to beamstrahlung never
causes the beam lifetime; the lifetime is always determined by
the emission of single photons.

In the ``crab waist'' collision scheme~\cite{crab,SuperB}, the beams
collide at an angle $\theta \gg \sigma_x/\sigma_z$.  The crab-waist
scheme allows for higher luminosity when it is restricted only by the
tune shift, characterized by the beam-beam strength parameter. One
should work at a beam-beam strength parameter smaller than some
threshold value, $\approx 0.15$ for high-energy SRCs~\cite{Zim}.

In head-on collisions, the vertical beam-beam strength parameter
(further ''beam-beam parameter'') ~\cite{Wied}
\begin{equation}
\xi_y =\frac{Nr_e\beta_y}{2\pi\gamma\sigma_x\sigma_y}\approx
\frac{Nr_e\sigma_z}{2\pi\gamma\sigma_x\sigma_y} \,\,\,\mbox{for}\,\,\, \beta_y \approx \sigma_z.
\label{xiy}
\end{equation}
In the crab-waist scheme~\cite{crab},
\begin{equation}
\xi_y =\frac{Nr_e\beta_y^2}{\pi\gamma\sigma_x\sigma_y\sigma_z}\,\,\,\mbox{for}\,\,\, \beta_y \approx \sigma_x/\theta.
\label{xiyc}
\end{equation}
The luminosity in head-on collisions
\begin{equation}
\lum \approx \frac{N^2f}{4\pi\sigma_x\sigma_y} \approx \frac{N f\gamma\xi_y }{2r_e\sigma_z};
\label{lhead}
\end{equation}
in crab-waist collisions,
\begin{equation}
\lum \approx \frac{N^2f}{2\pi\sigma_y\sigma_z \theta} \approx \frac{N^2 \beta_y f}{2\pi\sigma_x\sigma_y\sigma_z}  \approx \frac{N f \gamma\xi_y }{2r_e\beta_y}.
\label{lcrab}
\end{equation}
In the crab-waist scheme, one can make $\beta_y \ll \sigma_z$, which
enhances the luminosity by a factor of $\sigma_z/\beta_y$ compared to
head-on collisions. For example, at the proposed Italian SuperB
factory~\cite{SuperB} this enhancement would be a factor of 20--30.

Using Eqs.~\ref{xiy} and \ref{xiyc} and the
restriction in Eq.~\ref{11}, we find the minimum beam energy
when beamstrahlung becomes important. For
head-on collisions,
\begin{equation}
\gamma_\mathrm{min}=\left(\frac{0.1 \eta \alpha
  \sigma_z^2}{6\pi r_e\xi_y\sigma_y} \right)^{1/2} \propto
\frac{\sigma_z^{3/4}}{\xi_y^{1/2}\varepsilon_y^{1/4}};
\end{equation}
for crab-waist collisions,
\begin{equation}
\gamma_\mathrm{min}=\left(\frac{0.1 \eta \alpha
  \beta_y^2}{3\pi r_e\xi_y\sigma_y} \right)^{1/2} \propto
\frac{\beta_y^{3/4}}{\xi_y^{1/2}\varepsilon_y^{1/4}}.
\end{equation}
In the crab-waist scheme, beamstrahlung
becomes important at much lower energies because $\beta_y \ll
\sigma_z$. This can be understood from Eq.~\ref{xiyc}: smaller $\beta_y$
corresponds to denser beams, leading to a higher beamstrahlung rate.

Examples: a) SuperB~\cite{SuperB}: crab waist, $E_0=7\gev$,
$\sigma_y = 20\nm$, $\beta_y=0.2\mm$, $\xi_y=0.16$. Then,
$E_\mathrm{min}=29\gev$, i.e., beamstrahlung is not
important. b) The STR3 project (Table~\ref{Table1}): crab crossing,
$E_0=120\gev$, $\sigma_y = 50\nm $, $\beta_y=0.25\mm$, $\xi_y \sim
0.2$. Then, $E_\mathrm{min}=16.5\gev$, a factor 7 lower than $E_0$;
thus, beamstrahlung is very important. c) For projects STR1
and STR2: head-on, $E_0=120\gev$, $\sigma_y = 500\nm $, $\sigma_z=3\mm$,
$\xi_y \sim 0.15$; $E_\mathrm{min}=68 \gev$, beamstrahlung is
important.

We have shown that beamstrahlung restricts the maximum value of
$N/\sigma_x\sigma_z$ and becomes important at energies $E_0 \gtrsim 70
\gev$ for \epem storage rings with head-on collisions; when the
crab-waist scheme is employed, this changes to the more strict $E_0
\gtrsim 20 \gev$. All newly proposed projects listed in
Table~\ref{Table1} are affected as they have $E_0 \geq 120 \gev$.

Now, let us find the luminosity \lum when it is restricted both by
beam-beam strength parameter and beamstrahlung.
For head-on collisions,
\begin{equation}
\lum \approx \frac{(Nf)N}{4\pi\sigma_x\sigma_y}, \;
\xi_y \approx \frac{Nr_e\sigma_z}{2\pi\gamma\sigma_x\sigma_y}, \;
\frac{N}{\sigma_x \sigma_z} \equiv k \approx 0.1\eta\frac{\alpha}{3\gamma {r_e}^2}
\label{lhead-2}
\end{equation}
and $\sigma_y\approx \sqrt{\varepsilon_y \sigma_z}$.
This can be rewritten as
\begin{equation}
\lum \approx \frac{(Nf)k \sigma_z}{4\pi\sigma_y},
\;\;\;\;\; \xi_y \approx \frac{kr_e\sigma_z^2}{2\pi\gamma\sigma_y},\;\;\;\;\;\sigma_y\approx \sqrt{\varepsilon_y \sigma_z}.
\label{lhead-3}
\end{equation}
Thus, in the beamstrahlung-dominated regime the luminosity is
proportional to the bunch length, and its maximum value is determined by
the beam-beam strength parameter. Together, these equations give
\begin{equation}
\lum \approx \frac{Nf}{4\pi}\left(\frac{0.1
  \eta \alpha}{3}\right)^{2/3}\left(\frac{2\pi \xi_y}{\gamma r_e^5
  \varepsilon_y} \right)^{1/3},
\label{lhead-4}
\end{equation} \vspace{-4mm}
\begin{equation}
\sigma_{z,\mathrm{opt}}=\varepsilon_y^{1/3} \left(\frac{6\pi\gamma^2 r_e \xi_y}{0.1\eta\alpha} \right)^{2/3}.
\label{sz}
\end{equation}
Similarly, for the crab-waist collision scheme,
\begin{equation}
\lum \approx \frac{(Nf)N \beta_y
}{2\pi\sigma_x\sigma_y\sigma_z}, \;
 \xi_y \approx \frac{Nr_e\beta_y^2}{\pi\gamma\sigma_x\sigma_y\sigma_z},
\; \frac{N}{\sigma_x \sigma_z}\equiv k  \approx
0.1\eta\frac{\alpha}{3\gamma {r_e}^2}
\label{lcrab-2}
\end{equation}
and $\sigma_y\approx \sqrt{\varepsilon_y \beta_y}$.
Substituting, we obtain
\begin{equation}
\lum \approx \frac{(Nf)k \beta_y}{2\pi\sigma_y}, \;\;\;\; \frac{kr_e\beta_y^2}{\pi\gamma\sigma_y} \approx \xi_y,\;\;\;\; \sigma_y\approx \sqrt{\varepsilon_y \beta_y}.
\label{lcrab-3}
\end{equation}
The corresponding solutions are
\begin{equation}
\lum \approx \frac{Nf}{4\pi}\left(\frac{0.2
  \eta \alpha}{3}\right)^{2/3}\left(\frac{2\pi \xi_y}{\gamma r_e^5
  \varepsilon_y} \right)^{1/3},
\end{equation} \vspace{-4mm}
\begin{equation}\beta_{y,\mathrm{opt}}=\varepsilon_y^{1/3}
\left(\frac{3\pi\gamma^2 r_e \xi_y}{0.1\eta\alpha} \right)^{2/3}.
\end{equation}
We have obtained a very important result: {\sl in the
  beamstrahlung-dominated regime, the luminosities attainable in
  crab-waist and head-on collisions are practically the same.}  The
gain from using the crab-waist scheme is only a factor of $2^{2/3}
\sim 1$, contrary to the low-energy case, where the gain may be
greater than one order of magnitude. For this reason, from this point
on we will consider only the case of head-on collisions.

From the above considerations, one can find the ratio of the luminosities
with and without taking beamstrahlung into account: it is
equal to $\sigma_z/\sigma_{z,\mathrm{opt}}$ for head-on collisions and
$\beta_y/\beta_{y,\mathrm{opt}}$ for crab-waist collisions and scales
as $1/E_0^{4/3}$ for $\gamma > \gamma_\mathrm{min}$.
\begin{table*}[htb]
\caption{Realistically achievable luminosities and other beam parameters for the
projects listed in Table~\ref{Table1} at synchrotron-radiation power $P=100$ MW.
Only the parameters that differ from those in Table~\ref{Table1} are shown.
}
\vspace{3mm}
{\renewcommand{\arraystretch}{0.88}
\begin{ruledtabular}
\begin{tabular}{l c c c c c c c c c c c}
& LEP & LEP3 & DLEP & STR1 & STR2 & STR3 & STR4 & STR5 & STR6 \\[-1mm]
&&&&&&cr-w&cr-w&cr-w&cr-w \\  \hline \\[-2mm]
$2E_0$, \gev & 209 & 240 & 240 & 240 & 240 & 240 & 400 & 400 & 500 \\
Circumference,\km & 27& 27 & 53 & 40 & 60 & 40 & 40 & 60 & 80 \\
Bunches/beam &$\sim 2$ &$\sim 7$& 70 & 24 & 53 & 240 & 36 & 45 & 31 \\
$N,\;10^{11}$  & 33 & 5.9 & 2.35 & 3.9 & 4. & 0.4 & 0.34 & 0.6& 0.65  \\
$\sigma_{z}$, \mm  & 8.1 &8.1 & 5.7 & 6.9 & 6.9 & 3.4 & 6.7 & 7.8 &9.6  \\
$\sigma_y$, \mum  & 1.4 & 1.1 & 0.53 & 0.78 & 0.78 & 0.19 & 0.27 & 0.36 &0.35  \\
\lum, $10^{34}\cms$ & 0.47 & 0.31 & 0.89 & 0.55 & 0.83 & 1.1 & 0.12 & 0.16 & 0.087
\end{tabular}
\end{ruledtabular}
\vspace{-1mm}
}
\label{Table2}
\end{table*}
In practical units,
\begin{equation}
\frac{\sigma_{z,\mathrm{opt}}}{\mm} \approx
\frac{2\xi_y^{2/3}}{\eta^{2/3}} \left
(\frac{\varepsilon_y}{\nm}\right)^{1/3}
\left(\frac{E_0}{100\gev}\right)^{4/3}; \;\;
\frac{\beta_{y,\mathrm{opt}}}{\sigma_{z,\mathrm{opt}}} \approx 0.63.
\label{sz-pract}
 \end{equation}
For example, for $\xi_y=0.15$, $\eta=0.01$, $E_0=100 \gev$ and the
vertical emittances from Table~\ref{Table1} ($\varepsilon_y=0.01$ to 0.15\nm),
we get $\sigma_{z,\mathrm{opt}}=2.5$ to 6.4\mm.

According to Eq.~\ref{lhead-4}, the maximum luminosity at high-energy
SRCs with beamstrahlung taken into account
\begin{equation}
\lum \approx
h\frac{N^2f}{4\pi\sigma_x\sigma_y}=h\frac{Nf}{4\pi}\left(\frac{0.1
  \eta \alpha}{3}\right)^{2/3}\left(\frac{2\pi \xi_y}{\gamma r_e^5
  \varepsilon_y} \right)^{1/3},
\label{19}
\end{equation}
where $h$ is the hourglass loss factor,  $f=n_\mathrm{b} c/2\pi R$ is the collision rate,
$R$ the average ring radius, and $n_\mathrm{b}$  the number of
bunches in the beam.

The energy loss by one electron in a circular orbit of radius
 $R_b$~\cite{pdg} $\delta E= 4\pi e^2 \gamma^4/3R_b$, then the power
 radiated by the two beams in the ring
\begin{equation}
P=2 \delta E \frac{cNn_\mathrm{b}}{2\pi R}=\frac{4e^2\gamma^4cNn_\mathrm{b}}{3RR_\mathrm{b}}.
\label{22}
\end{equation}
Substituting $Nn_\mathrm{b}$ from Eq.~\ref{22} to Eq.~\ref{19}, we obtain
\begin{equation}
\lum \approx h\frac{(0.1\eta\alpha)^{2/3}P R}{32\pi^2\gamma^{13/3} r_e^3}
\left(\frac{R_\mathrm{b}}{R}\right)
\left(\frac{6\pi\xi_yr_e}{ \varepsilon_y} \right)^{1/3},
\label{23}
 \end{equation}
or, in practical units, \vspace{-2mm}
$$\frac{\lum}{10^{34}\cms} \approx \frac{100h\eta^{2/3} \xi_y^{1/3}}{(E_0/100\gev)^{13/3}
(\varepsilon_y/\nm)^{\frac{1}{3}} } $$ \vspace{-5mm}
\begin{equation}\;\;\;\;\;\;\;\;\; \;\;\;\;\;\times
\left(\frac{P}{100\, \mbox{MW}}\right)
\left(\frac{2\pi R}{100\km}\right) \frac{R_\mathrm{b}}{R} \vspace{-1mm}
\label{29}
\end{equation}
Once the vertical emittance is given as an input parameter, we find
the luminosity and the optimum bunch length by applying
Eq.~\ref{sz-pract}.  Beamstrahlung and the beam-beam strength
parameter determine only the combination $N/\sigma_x$; additional
technical arguments are needed to find $N$ and $\sigma_x$ separately.
When they are fixed, the optimal number of bunches $n_\mathrm{b}$ is
found from the total SR power, Eq.~\ref{22}.

In Table~\ref{Table2}, we present the luminosities and beam parameters
for the rings listed in Table~\ref{Table1} after beamstrahlung
is taken into account.  The following assumptions are made: SR power
$P=100$ MW, $R_\mathrm{b}/R=0.7$, $h=0.8$, $\xi_y=0.15$, $\eta=0.01$; 
the values of $\varepsilon_y$, $\varepsilon_x$ and $\beta_x$  are
taken from Table~\ref{Table1}.

Comparing Tables~\ref{Table1} and \ref{Table2}, one can see that at
$2E_o=240 \gev$ taking beamstrahlung into account lowers the
luminosities at storage-ring colliders with crab-waist collisions by a
factor of 15.  Nevertheless, these luminosities are comparable to
those at the ILC, $\lum_{\mathrm{ILC}} \approx (0.55$--$0.7) \times
10^{34} \cms$ at $2E_0=240 \gev$~\cite{ILCinterim}.  However, at
$2E_0=500 \gev$ the ILC can achieve $\lum_{\mathrm{ILC}} \approx
(1.5$--$2) \times 10^{34} \cms$, which is a factor 15--25 greater than
the luminosities achievable at storage rings.

In conclusion, we have shown that the beamstrahlung phenomenon must be
properly taken into account in the design and optimization of
high-energy \epem storage rings colliders (SRC). We have demonstrated
that beamstrahlung suppresses the luminosities as $1/E_0^{4/3}$ at
energies $E_0 \gtrsim 70 \gev$ for head-on collisions and $E_0 \gtrsim
20 \gev$ for crab-waist collisions.  Beamstrahlung makes the
luminosities attainable in head-on and crab-waist collisions
approximately equal above these threshold energies.  At $2E_0 =
240$--$500 \gev$, beamstrahlung lowers the luminosity of crab-waist
rings by a factor of 15--40. Some increase in SRC luminosities can be
achieved at rings with larger radius, larger energy acceptance, and
smaller beam vertical emittance.

We also conclude that the luminosities attainable at \epem storage
rings (at one interaction point) and linear colliders are comparable
at $2E_0 = 240 \gev$.  However, at $2E_0 = 500 \gev$ storage-ring
luminosities are smaller by a factor of 15--25. Linear colliders
remain the most promising instrument for energies $2E_0 \gtrsim
250 \gev$.

This work was supported by Russian Ministry of Education and Science.

\end{document}